\documentclass[conference]{IEEEtran}
\IEEEoverridecommandlockouts
\usepackage{silence}
\usepackage{cite}
\usepackage{amsmath,amssymb,amsfonts}
\usepackage{algorithmic}
\usepackage{graphicx}
\usepackage{textcomp}
\usepackage{xcolor}
\def\BibTeX{{\rm B\kern-.05em{\sc i\kern-.025em b}\kern-.08em
    T\kern-.1667em\lower.7ex\hbox{E}\kern-.125emX}}
\begin{document}

\title{Estimating the Effective Topics of Articles and journals Abstract Using LDA And K-Means Clustering Algorithm\\

}
\author{\IEEEauthorblockN{1\textsuperscript{st} Shadikur Rahman}
\IEEEauthorblockA{\textit{dept. of Software Engineering} \\
\textit{Daffodil International University }\\
Dhaka, Bangladesh \\
shadikur35-988@diu.edu.bd}
\and
\IEEEauthorblockN{2\textsuperscript{nd} Umme Ayman Koana}
\IEEEauthorblockA{\textit{dept. of Electronics \& Communication Engineering} \\
\textit{Khulna University of Engineering \& Technology}\\
Khulna, Bangladesh \\
koana2k12@gmail.com}
\and
\IEEEauthorblockN{3\textsuperscript{rd} Aras M. Ismael}
\IEEEauthorblockA{\textit{dept. of Information Technology} \\
\textit{Sulaimani Polytechnic University}\\
Kurdistan, Iraq \\
Aras.masoud@usa.com}
\and
\IEEEauthorblockN{4\textsuperscript{th} Karmand Hussein Abdalla}
\IEEEauthorblockA{\textit{dept. of Software Engineering} \\
\textit{University of Raparin }\\
Ranya, Iraq \\
karmand.hussein@uor.edu.krd}

}

\maketitle
\thispagestyle{plain}
\pagestyle{plain}

\begin{abstract}
Analyzing journals and articles abstract text or documents
using topic modeling and text clustering becomes modern solutions for the increasing number of text documents.Topic modeling and text clustering are both intensely involved tasks that can benefit one another. Text clustering and topic modeling algorithms are used to maintain massive amounts of text documents. In this study, we have used LDA, K-Means cluster and also lexical database WordNet for keyphrases extraction in our text documents. K-Means cluster and LDA algorithms achieve the most reliable performance for Keyphrases extraction in our text documents. This study will help the researcher to make searching string based on journals and articles by avoiding misunderstandings.
\end{abstract}


\section{Introduction}
Topic modeling and text clustering perform very significant roles in many research areas, such as text mining, text labels, natural language processing, text classification, and information retrieval.Recently, The ever-increasing electronic documents which is becoming a challenging task to extract out the exact topics without reading the entire text documents. It provides a researcher or reader an advantage to find out what is going on through the journals and articles text documents. However, for multidimensional issues, it is a bit tedious for the reader or researcher to understand the point. If it is labeled correctly, it is better to understand it.

There are various algorithms for topic modeling and text clustering. The most popular are LDA~\cite{b3}, LSA~\cite{b5}, NMF~\cite{b11}, pLSA \cite{b7}, and HDP \cite{b17} and K-Means cluster~\cite{b12}.We have chosen popular topic modeling and text clustering technique i.e. Latent Dirichlet Allocation (LDA) and K-Means cluster respectively are handled to extract latent topics and clusters of documents. These techniques can automatically identify the abstract topics that occur in a collection of text documents. Topic modeling and text clustering are two widely studied problems that have many applications.Text clustering strives to organize similar documents into groups, which is essential for text document organization, text labels, summarization, classiﬁcation, and retrieval. Topic modeling develops probabilistic generative models to discover the latent semantics embedded in document collection and has illustrated widespread success in modeling and analyzing text documents. 

In our previous studies~\cite{b8} and~\cite{b16}, we proposed a model to ﬁnd generic labels for the polynomial topics over short text documents. In which we have used only LDA model to generate topic models. And also proposed to find out automated labels with the description for the perennial topics in short online text documents. we have used the same model for generating the generic label and used LDA, LSI, and NMF for training the model. We have operated these models over the short text documents and measure the WUP similarities of each label. By comparing the WUP similarities for each model, we come to the conclusion that we can ﬁnd the exact labels by using the LDA model. Though we have done this experiment in short text documents, it can be done on large texts or documents. Correspondingly,  We have used a unique technique of topic modeling and text clustering to solve text document problems for automated labeling.  We have used LDA, and K-Means cluster for training the model and also used Word2Vec~\cite{b6} to classify the vectors of similar words together in vector space. We have operated these models over the large scale journals and articles text documents and measure the WUP similarities of each label. By comparing the WUP similarities for each model, We have come to the conclusion that using the technology we can label properly.

The rest of the paper is organized as follows: Related Work is discussed at section~\ref{Related Work} followed by Research Methodology and Result \& Discussion at Section~\ref{Research Methodology} and~\ref{Result & Discussion} respectively. And finally Section~\ref{Conclusion} concludes the summary.

\section{Related Work}\label{Related Work}
The efficiency of topic labeling to automatically assigning labels to topics is described by~\cite{b14} for LDA topics which were an unsupervised approach.In their approach, they showed potential ways to automatically label multinomial models in objective ways.In~\cite{b16} used the topic model to create generic labels and used LDA, LSI and NMF to train the model.They conducted this procedure on short text documents and measured the WUP similarity of each label. By comparing the WUP matches for each model we conclude that we can make the correct labels using the LDA model.In this~\cite{b19} proposed multi-grain clustering topic model (MGCTM) which integrates text clustering and topic modeling into together performs the two tasks to achieve the overall best performance. In \cite{b9} presented an approach for topic labeling. Primary, they created a set of candidate labels from the top-ranking topic terms, titles of Wikipedia articles including the top-ranking topic terms, and sub phrases extracted from the Wikipedia article titles. Then ﬁnally they ranked the label candidates using a sequence of association measures, lexical features, and an information retrieval feature. Natural language processing(NLP) is also used for topic modeling and labeling method.Using Word2Vec word embedding, \cite{b1} labeled online news considering the article terms and comments.In this work,~\cite{b13} proposed a method for labeling topics induced by a hierarchical topic model. Their label candidate set is the Google Directory service (gDir) hierarchy, and label selection takes the form of ontological alignment with gDir. In recent work,~\cite{b10} proposed to approach topic labeling via best term selection. Selecting one of the top ten topic terms to label the overall topic. In \cite{b2} author presented an approach to select the most relevant labels for topics by computing neural embedding for documents and words. 

The phrase-based labels in the above works are still pretty short and are sometimes not sufficient for interpreting the topics. Though several approaches had been proposed in various studies, as per our knowledge no proof is present for assessing the models for generating topic labels.

\section{Research Methodology}\label{Research Methodology}
\begin{figure}
\includegraphics[width=8.7cm, height=4cm]{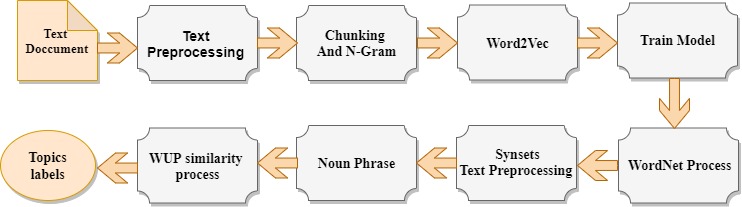}
\caption{Research Methodology}\label{research_methodology}
\end{figure}

In this section, the entire process of our research activities has been described. In the first instance, we have selected our journals and articles datasets~\footnote{https://github.com/sadirahman/Estimating-the-Effective-Topics-of-Articles-and-journals/tree/master/Datasets}. For the dataset, we have selected some online journals and articles documents to perform our research process. As preparation, We have to take several steps to complete the process such as text pre-processing, Chunking and N-gram, Word2Vec, training model, WordNet Process, noun phrase selection, and label processing with the help of WordNet Synset. Then we receive topic labeling based on our topic modeling and text clustering. We transfer a retrieve responsibility to compare three topic representations: (1) Clustering result, (2) Topic labels and (3) Keyphrases Extraction. Figure~\ref{research_methodology} shows an overview of our research experiment.

\subsection{\textbf{Text Preprossessing}}
\begin{figure}
\includegraphics[width=8.7cm, height=2cm]{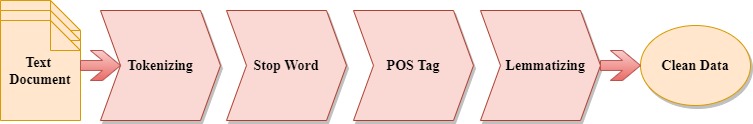}
\caption{Text Preprossessing process}\label{Preprossessing}
\end{figure}

To analyze our text document datasets, we preprocess unsupervised journals and articles text documents. In this research, we have used online journals and articles. In most of the cases, we realize that text data is not effectively cleaned. For cleaning the text data, text pre-processing is required. For doing the pre-processing, we need to follow several steps like tokenization, stop words removing,  POS tag, lemmatizing and removing punctuation.  Figure~\ref{Preprossessing} shows an overview of our text document prepossessing process.

\paragraph{\textbf{Text Tokenization process}}
We create a bag of words in our text documents and divide them into punctuation marks. However, we make sure that small words such as “don't”, “I'll”, “I'd” remain as one word. Splitting the provided text into smaller portions called a token.
\paragraph{\textbf{Stop Words Removing}}
Stop words are common words in a language like "almost", "an", "the", "is", "in". These words have no significant meaning, and we usually remove words from our text documents.

\paragraph{\textbf{POS Tag process}}
We used the NLTK word tokenizer to parse each text into a list of words. After that Text Tokenization, then we used Pos tagging in NLP using NLTK. The part of speech(POS tag) explains the corpus of how a word is used in a sentence.POS tag is separated into subclasses. POS Tagging solely means labeling words with their appropriate Part-Of-Speech. There are eight main parts of speech - nouns, pronouns, adjectives, verbs, adverbs, prepositions, conjunctions, and interjections.

\paragraph{\textbf{Lemmatizing process}}
We used the Lemmatizing process of decreasing words to their word Lemma, base or root form, for example, worked–work, loved–love, roads-road, etc.

\subsection{\textbf{Chunking And N-Gram process}}
After Preprossessing, We use the N-gram and chunking process to sort the sentence sequence of N words and extracting phrases from our text documents.

\begin{figure}
\includegraphics[width=8.7cm, height=5cm]{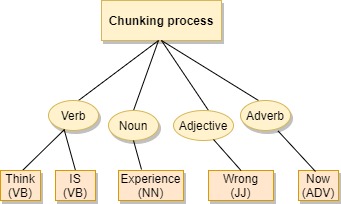}
\caption{Overview of Chunking Process}\label{chunk}
\end{figure}

\paragraph{\textbf{Chunking process}}
We used the chunking process of extracting phrases from unstructured text documents. Chunking works on top of POS tagging, it uses POS tags as input and provides chunks as output. We search for chunks corresponding to an individual some POS tag phrase. The journals and articles include many parts of speech that are irrelevant to detect semantic orientation in our case. We consider only Adjectives (JJ), Verb (VB), Adverb (ADV) and Noun (NN) Parts of Speech (POS) tag from Penn Treebank annotation. For example, ”I think the experience is wrong now”.Figure~\ref{chunk} shows an overview of our chunking process.

\paragraph{\textbf{N-Gram Model process}}
An N-gram is a sequence of N words, which computes p(w—h), the probability of a word w*~\cite{b4}. We have used the N-gram model using the journals and articles text documents held-out data as we used for the word-based Natural language process model we discussed above in our research. The purpose of using this is to maintain the sequence of candidate labels in our text document. For example, if we consider the text- ”Please enter your abstract text”. For this text, the 2-gram and 3-gram words are presented in Table~\ref{n-gram}.

\begin{table}[hbt!]
\centering
\setlength{\tabcolsep}{20pt}
\caption{Process of an N-gram}\label{n-gram}
\begin{tabular}{c c}
\hline
2-gram(bigram) & 3-gram(trigram)  \\
\hline
"Please enter" &"Please enter your"	\\
"Enter your" & "Enter Your abstract"\\
"Your abstract" & "Your abstract text" \\
"Abstract text" & ------\\

\hline
\end{tabular}
\end{table}

\subsection{\textbf{Word2Vec Model Process}}
After the previous process, we used the word2vec method in our research work. Word2vec gives direct access to vector representations of journals and articles text words. Word2vec~\cite{b6} is to classify the vectors of similar words together in vector space. It recognizes similarities mathematically. Word2vec produces vectors that are distributed numerical representations of word features, features such as the context of individual words. Word2Vec Model based on journals and articles text documents and gensim word2vec module. We used word2vec for the most similar word-ﬁnding parameter ”topn = 10”.For example, the word ”Cluster” is a vector representation in Table~\ref{wordvec} 

\begin{table}[hbt!]
\centering
\setlength{\tabcolsep}{8pt}
\caption{Process of Word2vec}\label{wordvec}
\begin{tabular}{l c}
\hline
Similar Words(Cluster) & Numerical Representations  \\
\hline
Corruption & 0.82 \\
Wheel & 0.82 \\
Weapons & 0.82 \\
Ministration & 0.81 \\
Blast & 0.81 \\
Sinners & 0.80 \\
Covert & 0.80 \\
Bird & 0.80 \\
Necessity & 0.80 \\
Transgression & 0.79 \\
\hline
\end{tabular}
\end{table}

\subsection{\textbf{Train model}}
\begin{figure}
\includegraphics[width=8.7cm, height=12cm]{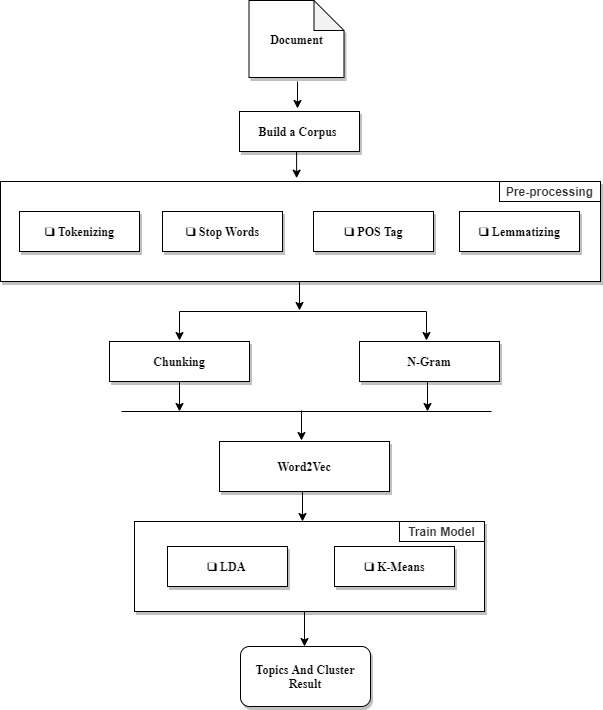}
\caption{Train Model}\label{Train_model}
\end{figure}

In this research, we used the most popular text clustering and topic modeling algorithms K-Means cluster, and LDA. In order to train our models, we must first need journal and article documents. Each topic model and text cluster is based on the same basic role: each document contains a mixture of topics, and each cluster is composed of a set of words. As a result, the purpose of the topic modeling and text clustering algorithms is to uncover these latent topics and cluster variables that shape the meaning of our text documents and training text documents. Figure~\ref{Train_model} shows an overview of our research training models.

\begin{figure}
\includegraphics[width=8.7cm, height=4cm]{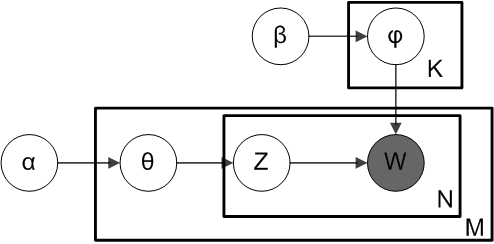}
\caption{Process of LDA model}\label{LDA_model}
\end{figure}

\paragraph{\textbf{Latent Dirichlet allocation(LDA)}}
Latent Dirichlet Analysis is a probabilistic model, and to obtain cluster assignments. LDA utilizes Dirichlet priors for the document-topic and word-topic distributions and it utilizes two probability values: P (word — topics) and P (topics — documents) \cite{b3}.Figure~\ref{LDA_model} shows an overview of our LDA models where K denotes the number of topics, M denotes the number of documents, N is number of words in a given document and w is the specific word.

To perform the LDA model works by deﬁning the number of ’topics’ that are begun in our set of training text documents. Now we will present the model output below: here we have chosen the number of topics = 3 and Number of Words = 3. We have also set the random state = 1 which is enough for our journals and articles text documents. Table~\ref{LDA} presents the result of training LDA model after executing Document S1.

\begin{table}[hbt!]
\centering
\setlength{\tabcolsep}{8pt}
\caption{Result of an example Document S1 in LDA model}\label{LDA}
\begin{tabular}{c c c}
\hline
Topic 1 & Topic 2 & Topic 3 \\
\hline
0.071*"topic"  & 0.048*"nmf" & 0.065*"document" \\
0.046*"algorithm"  & 0.047*"used" &	0.045*"topic" \\
0.034*"document" &	0.046*"lsi" & 0.037*"modeling"\\
\hline
\end{tabular}
\end{table}

\begin{figure}
\includegraphics[width=8.7cm, height=8cm]{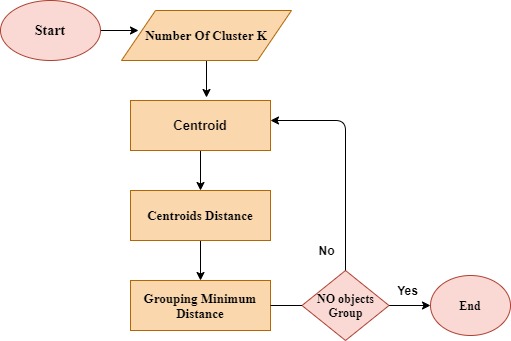}
\caption{Process Flow of K-Means Cluster}\label{Kmeans}
\end{figure}

\paragraph{\textbf{K-Means Cluster}}
K-means clustering~\cite{b12} is a method usually used to automatically partition a data set into k groups. It is an unsupervised learning algorithm.k-means is to minimize the total sum of the squared distance of every point to its corresponding cluster centroid. K-Means clustering aims to partition n observations into k clusters in which each observation belongs to the cluster with the nearest mean, serving as a prototype of the cluster.Figure~\ref{Kmeans} shows process flow of K-means Clustering algorithm.

We train the clustering algorithm for over our text document to split it into word groups. To perform the K-Means cluster algorithm begins by defining the number of ’clusters’ that is begun in our set of training text documents. Now we will present the model output below:  here we have chosen the number  K = 3  and  number of words = 3.  We have also set the n-init = 1 and max-iter = 100 which is enough for  our journals and  articles  text documents.  Table~\ref{KMC} presents  the  result  of  training K-Means cluster algorithm after executing Document S1.

\begin{table}[hbt!]
\centering
\setlength{\tabcolsep}{20pt}
\caption{Result of an example Document S1 in K-Means cluster}\label{KMC}
\begin{tabular}{c c c}
\hline
Cluster 1 & Cluster 2 & Cluster 3 \\
\hline
Lda  & Documents & Topics \\
Nmf & Popular &	Meaning \\
Used & Lsi & Select \\
\hline
\end{tabular}
\end{table}

\subsection{\textbf{WordNet process}}
\begin{figure}
\includegraphics[width=8.5cm, height=12cm]{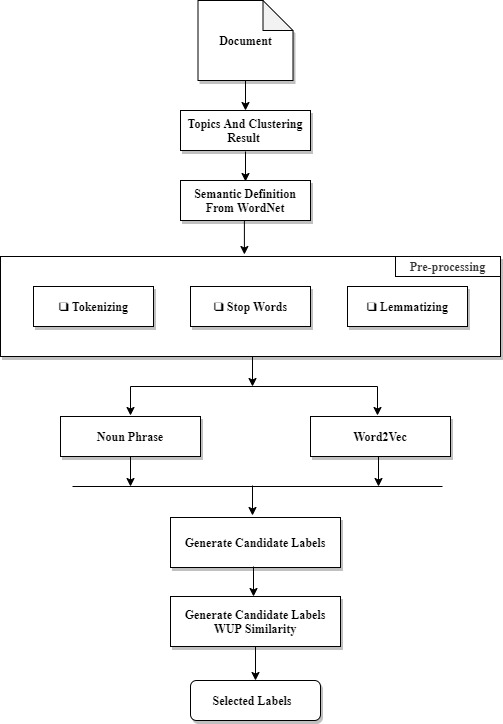}
\caption{Process Flow of WordNet process}\label{wordnet}
\end{figure}

WordNet~\cite{b15} is a comprehensive lexical database of English. Nouns, verbs, adjectives, and adverbs are classiﬁed into sets of cognitional synonyms (synsets), each meaning a distinct idea. It partly compares a dictionary, in that it classiﬁes words respectively based on their suggestions. However, there are any signiﬁcant variations. First, it interlinks not just word makes sequences of words but special functions of words. As a result, words that are seen in near concurrence to one extra in the system are semantically disambiguated. Second, it speciﬁes the semantic relationships between words, whereas the classiﬁcation of words in a dictionary does not match any speciﬁc design other than discovering the identity. For doing the WordNet process, we need to follow several steps such as synsets text preprocessing, Noun Phrase process and WUP similarity process. Figure~\ref{wordnet} shows an overview of our text document WordNet process for keyphrases extraction.

In our topic modeling and text clustering result in our most comprehensive relevant words and next works in WordNet term. This WordNet term gives a word deﬁnition in our selected top-weighted word. Suppose our top-weighted selected word is “algorithm” then WordNet synset gives a deﬁnition are: "S: (n) algorithm, algorithmic rule, algorithmic program (a precise rule (or set of rules) specifying how to solve some problem)". Then we started again preprocessing in our WordNet deﬁnition and also pick up the noun and proper noun phrase.

\paragraph{\textbf{Noun Phrase process}}
After synsets text preprocessing, we have only picked the noun and proper noun from the preprocessed result. Applying this approach, the topic is taken by top nouns words with the largest frequency in the text corpus. For noun phrase choosing: ﬁrst, the tokenization of text is executed to lemma out the words. The tokenized text is then tagged with parts of speech NN (nouns), NNP (proper nouns), VB (verbs), JJ (adjectives), etc. Before lemmatizing and stop-words removing, parts-of-Speech (POS) tagging is done. The stop-words are removed after POS tagging. In the ﬁnal stage, words including their tags and rounds are put in a hash table and most solid nouns are obtained from these to create a heading for a text. The results of noun phrase words are presented in Table~\ref{noun_phrase}.

\begin{table}[hbt!]
\centering
\setlength{\tabcolsep}{8pt}
\caption{Result of an Noun Phrase words}\label{noun_phrase}
\begin{tabular}{c c}
\hline
POS tagging words & Only Noun words  \\
\hline
LDA, K-Means , Used , & LDA, K-Means, Topic, \\
Topic, Select, Transportation & Transportation \\

\hline
\end{tabular}
\end{table}

\paragraph{\textbf{WUP similarity Labels process}}
After the Noun phrase process, we take topics and clusters from the documents we only get top-weighted words from the topic and cluster because its result is well for its topic set and cluster set. Then we explore the semantic deﬁnition from the lexical database for English parts of speeches which are termed WordNet. We choose the drfination description because within its phrases we find the candidate labels. After pre-processing the deﬁnition, we prepare the candidate labels and from those candidate labels, we estimate the candidate labels with the main topic word for conceptual semantic relatedness measure by WUP similarity~\cite{b18}. Then we begin in our WUP similarity process for labeling. The results of WUP similarity Labels process  are presented in Table~\ref{WUP}.

\begin{table}[hbt!]
\centering
\setlength{\tabcolsep}{1pt}
\caption{WUP similarity Labels process }\label{WUP}
\begin{tabular}{c|l|l|c|c}
\hline
Document (S1) & Words & Top weighted  & Candidate Labels  & Labels\\
&& word(s) &\\ \hline
&Topic &  & Matter,Conversation,&\\
Topic 1 &Algorithm & Topic & Discussion, Situation,& Conversation \\
&Document &	 & Event&\\\hline

\end{tabular}
\end{table}

\section{\textbf{Result \& discussion}}\label{Result & Discussion}
We estimate the text clustering and topic modeling performance of our model based on journals and articles text documents. In this section, we consider the overall results of our research work. We have picked the top three words for each model and obtained the top word considering the highest weighted value. Then we obtain the definition of top weighed word using the WordNet synsets. After that, we pre-process the found definition and choose the candidate labels of each definition by using Noun-phrase and Word2Vec.

Finally, we estimate the Keyphraes (label) comparing the WUP similarities accuracy between each candidate label and the top-weighted word. Table~\ref{label_LDA} and Table~\ref{label_Kmeans} are showing the details of selecting the labels of three journals and articles text documents by using topic modeling algorithm LDA and Text clustering algorithm K-Means respectively.

\begin{table*}[hbt!]
\centering
\setlength{\tabcolsep}{7pt}
\caption{Label of chosen 3 documents with LDA}\label{label_LDA}
\begin{tabular}{c|l|l|l|l|l}
\hline
Document(s) & Topic(s) & Topic  & Top Weighted  & Candidate  & Label \\ 
& & Word(s) & Word(s) & Label(s) &\\
\hline
S1  &	Topic 1	 & Topic, Algorithm,  & Topic & Matter, Conversation,  & Conversation \\ & & Document & & Discussion, Situation, Event &\\

& Topic 2 & Use,Nmf,Lsi & Use & Employ,Service,Purpose, & Service \\ & & & & Act,Put &\\

& Topic 3 & Document, Topic, & Document & Represtation & Information\\
&& Modeling & & Information,Instruction,\\
&&&& File,Obligation & \\ \hline

S2 & Topic 1& News
& News & Information,Event & Information	\\
&& Misconception &  & Magazine,Newspaper &\\
&& Avoiding && Commentary &\\
& Topic 2 &	News, Headline, 	& News&	Information,Event  &	Information\\
&& Increasing &&Magazine,Newspaper & \\
&&&& Commentary &\\
& Topic 3 & Headline,News & Headline & Caption,Newspaper 	& Story\\
&& Sentiment & & Action,Story,page &\\\hline

S3 & Topic 1 &	Topic,word,   &	Topic &	Matter,Conversation 	& Conversation\\
&& Reproduced && Discussion,Situation,Event &\\
& Topic 2 & Label,Topic & Label	& Identification,Description,  &	Identification\\
&&  labeling && Name,Mechanism &\\
&&&& Reaction &\\

&Topic 3 &	Topic,Labeling, &	Topic &	Matter,Conversation  &	Conversation \\
&& Idea && Discussion,Situation,Event &\\\hline

\hline
\end{tabular}
\end{table*}

\begin{table*}[hbt!]
\centering
\setlength{\tabcolsep}{7pt}
\caption{Label of chosen 3 documents with K-means cluster}\label{label_Kmeans}
\begin{tabular}{c|l|l|l|l|l}
\hline
Document(s) & Cluster(s) & Cluster  & Top Weighted  & Candidate  & Label \\ 
& & Word(s) & Word(s) & Label(s) &\\
\hline
S1  &	Cluster 1	 & Topic, Study,  & Topic & Matter, Conversation,  & Conversation \\ & & Labelling & & Discussion, Situation, Event &\\

& Cluster 2 & Documents & Document & Representation, & Information \\ & &Popular,Everyday & & Information,Instruction \\
&&&& File,Obligation &\\

& Cluster 3 & Algorithm,Lda, & Algorithms & Rule,Solve,Problem,Set & Rule\\
&& Topic  & \\ \hline

S2 & Cluster 1& News,Headlines
& News & Information,Event & Information	\\
&& Interactivity &  & Magazine,Newspaper, &\\
&&&& Commentary &\\
& Cluster 2 &	News, Sentiment, 	& News&	Information,Event  &	Information\\
&& Headlines &&Magazine,Newspaper, & \\
&&&& Commentary &\\
& Cluster 3 & Sentiment, & Sentiment & Felling,Emotion 	& Felling(0.90)\\
&& Accuracy,Achieves & & Judgement,Tender,Belief &\\\hline

S3 & Cluster 1 &	Topic,Labels,   &	Topic &	Matter,Conversation 	& Conversation\\
&& Words && Discussion,Situation,Event &\\
& Cluster 2 & User,Cognitional & User	& Person,Someone,use  &	Person\\
&&  Understanding && Thing,Drug &\\

& Cluster 3 &	Noun,Phrase, &	Noun &	Content,Person,Word,  &	Preposition \\
&& Proposed && Preposition,Object &\\\hline

\hline
\end{tabular}
\end{table*}

Here choosing the ﬁnal label of each topic and cluster, WUP similarities values are used. The values of WUP similarities accuracy are showing in Table~\ref{WUP_LDA} and Table~\ref{WUP_kmeans} for models LDA and K-Means cluster respectively.

\begin{table}[hbt!]
\centering
\setlength{\tabcolsep}{2pt}
\caption{WUP similarity between topic and label with LDA}\label{WUP_LDA}
\begin{tabular}{c|l|l|c|c}
\hline
Document(s) & Top Weighted & Labels  & WUP similarity  & Average WUP\\
& word(s) &\\ \hline
&Topic & Conversation & 0.54&\\
S1 & Use &	Service &	0.80 & 0.62\\
&Document &	Information & 0.54&\\\hline
&News&	Information &	0.90&\\
S2 & News	& Information & 0.90 & 0.73\\
&Headline &	Story &	0.40&\\\hline

&Topic &	Conversation &	0.54&\\
S3 &Label &	Identification &	0.62& 0.56\\
&Topic &	Conversation & 0.54 &\\\hline

\end{tabular}
\end{table}

\begin{table}[hbt!]
\centering
\setlength{\tabcolsep}{2pt}
\caption{WUP similarity between cluster and label with K-Means }\label{WUP_kmeans}
\begin{tabular}{c|l|l|c|c}
\hline
Document(s) & Top Weighted & Labels  & WUP similarity  & Average WUP\\
& word(s) &\\ \hline
&Topic & Conversation & 0.54&\\
S1 &Document &	Information & 0.54& 0.47\\
&Algorithm &	Rule & 0.35&\\\hline
&News&	Information &	0.90&\\
S2 & News	& Information & 0.90 & 0.90\\
&Sentiment &	Feeling &	0.90&\\\hline

&Topic &	Conversation &	0.54&\\
S3 &User &	Person &	0.80& 0.69\\
&Noun &	Preposition & 0.75 &\\\hline

\end{tabular}
\end{table}

After choosing the topic and cluster labels,  then getting the WUP similarities values of each label, we average the WU similarities values. We can see in Table~\ref{WUP_comparision} that LDA model shows the accuracy (63\%) and the k-Means cluster models shows the accuracy (68\%). Both our models work together to give us the best results in our keyphrases extraction based on our journals and articles text documents.

\begin{table}[hbt!]
\centering
\setlength{\tabcolsep}{7pt}
\caption{WUP similarities difference among models}\label{WUP_comparision}
\begin{tabular}{c|l|c|c}
\hline
Models & Document Sets & Average WUP & Total Average\\\hline
&Document S1 &	0.62&\\
LDA &Document S2 &	0.73 & 0.63\\
&Document S3	& 0.56&\\\hline
 
&Document S1	& 0.47&\\
K-Means & Document S2	& 0.90 & 0.68\\
&Document S3 &0.69&\\\hline

\end{tabular}
\end{table}

\section{\textbf{Conclusion}}\label{Conclusion}
In this research, We presented a unique technique of topic modeling and text clustering to solve text document problems for automated labeling. This technique observed that our proposed method is an application to ﬁnd the relevant topic and cluster label for the polynomial topic in the text documents of our journals and articles. We have used topic modeling LDA and text clustering K-Means algorithm to train our model. For each model, we have determined the top three words and picked the top-weighted word to obtain the definition of chosen top words and generate the candidate labels for each of the words using the WordNet synset of the lexical database. We have compared the WUP similarities between candidate labels and top words, we chose the most accurate label for topics. This research can be broadly expanded into topics and labels for large scale text documents and a company text document can be effectively solved the problem.

\end{document}